# Distributed Optimization Strategy for Multi Area Economic Dispatch Based on Electro Search Optimization Algorithm

Mina Yazdandoost, *Member, IEEE,* Peyman Khazaei, *Student Member, IEEE,* Salar Saadatian, *Student Member, IEEE,* Rahim Kamali, *Student Member, IEEE*

*Abstract*—A new adopted evolutionary algorithm is presented in this paper to solve the non-smooth, non-convex and non-linear multi-area economic dispatch (MAED). MAED includes some areas which contains its own power generation and loads. By transmitting the power from the area with lower cost to the area with higher cost, the total cost function can be minimized greatly. The tie line capacity, multi-fuel generator and the prohibited operating zones are satisfied in this study. In addition, a new algorithm based on electro search optimization algorithm (ESOA) is proposed to solve the MAED optimization problem with considering all the constraints. In ESOA algorithm all probable moving states for individuals to get away from or move towards the worst or best solution needs to be considered. To evaluate the performance of the ESOA algorithm, the algorithm is applied to both the original economic dispatch with 40 generator systems and the multi-area economic dispatch with 3 different systems such as: 6 generators in 2 areas; and 40 generators in 4 areas. It can be concluded that, ESOA algorithm is more accurate and robust in comparison with other methods.

## Nomenclature

*Indices*

| | |
|---|---|
| $i, j$ | generating unit indices |
| $k$ | Iteration index |
| $l$ | Prohibited operating zone index |
| $t$ | Candidate solution index |
| $M$ | Number of area |

*Constants*

| | |
|---|---|
| $a_{ij}, b_{ij}, c_{ij}, e_{ij}, f_{ij}$ | cost coefficients of $jth$ generator in the $ith$ area |
| $B^i_{qj}$ | loss coefficient relating the productions of $qth$ and $jth$ generators in area $i$ |
| $B^i_{0j}$ | loss coefficient associated with the production of $jth$ generator in area $i$ |
| $B^i_{00}$ | loss coefficient parameter (MW) in area $i$ |
| $L_i$ | number of POZs for ith generator |
| $N_g$ | number of generating units |
| $P_{gi\min}$ | lowest output power of ith generator (MW) |
| $P_{gi\max}$ | highest output power of ith generator (MW) |
| $P_{gi,l}^{Low}, P_{gi,l}^{Up}$ | minimum/maximum boundary of the lth POZ for ith generator, respectively |
| $rand(1,n)$ | $(1 \times n)$ vector consists of random numbers in the range [0,1] |

*Variables*

| | |
|---|---|
| $F(pg)$ | generating unit cost function |
| $H(X)$ | objective function |
| $P_{gi}$ | power output of generating unit i (MW) |
| $T_{i,j}$ | Transmission power between area i and j areas |

## I. Introduction

ECONOMIC dispatch is highly concerned when optimization and power system are the topic of discussion. Allocating the required power among the committed generators and minimizing the cost function is the main aim of the economic dispatch. Moreover, it is demanded to satisfy all the physical and operational constraints at the mean-time [1]. The original economic dispatch problem is a

Mina Yazdandoost is with the Jahan Pardazesh Alborz Engineering Company, Karaj, Iran (email:mina.yazdandoost@yahoo.com)

Peyman Khazaei is with the Department of Electrical and Computer Engineering, Shiraz University of Technology, Shiraz, Iran (email:mina. electronic.peyman@gmail.com)

S. Saadatian is with the Department of Mechanical, Louisiana State University, Baton Rouge, USA. (e-mail: ssaada1@lsu.edu).

R. Kamali is with the Department of Electrical and Computer Engineering, Shiraz University of Technology, Shiraz, Iran. (e-mail: kamali.rahim@gmail.com).

second order polynomial problem. However, for modeling the valve point loading effect a sinusoidal term should be added [2].

Based on the literature, the economic dispatch problem is solved by different mathematical technique such as Lambda Iteration [3], Gradient Method [4] and linear programming [5]. However, using mathematical methods are not so suitable because of the discontinuity and nonlinearity of the fuel cost which is resulted from the valve point effect [6]. Although in some researches, dynamic programming was utilized for economic dispatch [2], this method is not recommended due to the curse of dimensionality [2]. Moreover, some meta-heuristics methods such as genetic algorithm (GA) [7,8], particle swarm optimization (PSO) [9-11], tabu search (TS) [12], simulated annealing (SA) [13], quasi-oppositional group search optimization (QOGSO) [14], chaotic global best artificial bee colony (CGABC) [15], Firefly algorithm (FA) [16], continuous quick group search optimizer (CQGSO) [17], fuzzy adaptive chaotic ant swarm optimization (FCASO) [18], augmented Lagrange hopfield network (ALHN) [19] were employed. It should be noted that, some of the mentioned methods do not assure the best and optimal solution.

Multi area economic dispatch (MAED) is an extension of economic dispatch problem [1]. The economic dispatch problem needs to be solved in each area and power exchange among the areas needs be determined whereas all constraints are satisfied and the cost function is minimized at the meantime. Furthermore, the transmission tie line capacity should be highlighted as one of the most significant distinctive features constraint in MAED. MAED has also been solved by using some mathematical methods such as linear programming [20], Dantzig–Wolfe decomposition principle [21] and decomposition approach using expert systems [22]. The mathematical problem will be more complex by increasing the problem's decision variables, the non-linearity because of the valve point effect, the discontinuity because of the prohibited operating zones and etc. Hence, some meta-heuristics methods such as particle swarm optimization (PSO) with Reserve-constrained multi-area environmental/economic dispatch [23], a new nonlinear optimization neural network approach [24], artificial bee colony optimization [25], teaching-learning based optimization (TLBO) in [1] and chaotic global best artificial bee colony [15] are proposed.

In this study, a new modified electro search optimization algorithm (ESOA) is utilized for solving the MAED is proposed. ESOA is proposed in [21] in power system operation and planning in which individual trying to approaches the best solution and far from the worst solution to get the best solution. This algorithm is simple, but, by increasing the dimensions of the problem, the performance of the algorithm reduces and defining the optimal solution is not guaranteed. Therefore, by considering all the possible states (such as individuals motion towards the worst solution or the individuals get away from the best solution or others states). In addition, a mutation operator is considered to increase the speed on convergence. Finally, the proposed method is applied for the multi-area economic dispatch problem under different conditions. Meta-heuristics methods are one of the successful tool to solve problem which is used in many areas and fields [26-29].

## II. Multi Area Economic Dispatch

### A. Original Economic Dispatch

Economic dispatch is considered as one of the most important optimization problems in power system field. The main goal of this problem is to define the generation level so that minimizes the cost while all constrained are satisfied. In order to model the economic dispatch, a quadratic function is utilized. However, in the huge generators the valve point effect and prohibited operating zones can lead to non-linearity and non-convexity of the cost function. Thus, for modeling the huge generators, a sinusoidal term needs to be added to the cost function as the valve point effect:

$$Min\ H(X) = \sum_{i=1}^{N_g} F_i(p_{gi})$$

$$F_i(p_{gi}) = a_i \times p_{gi}^2 + b_i \times p_{gi} + c_i + \left| e_i \times sin(f_i \times (p_{gi\ min} - p_{gi})) \right| \quad (1)$$

Where,

$$Z = [p_{g1}, p_{g2}, p_{g3}, \ldots, p_{gN_g}] \quad (2)$$

Figure. 1 demonstrates the power generator fuel curve with and without the valve point effect.

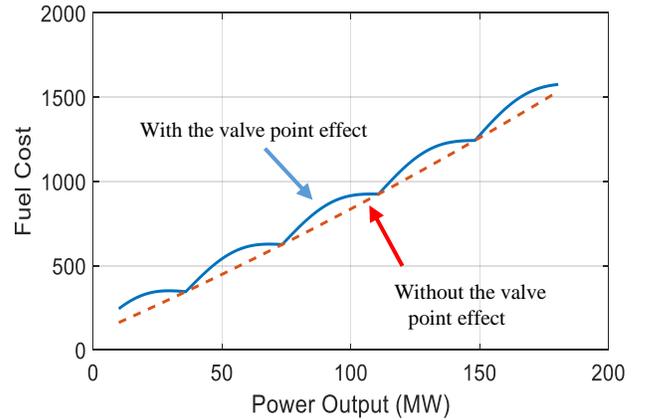

Fig. 1, The cost function with and without the valve point effect

### B. Multi-area Economic Dispatch

Multi area economic dispatch (MAED) is an advanced economic dispatch problem. The aim of MAED is to determine the power generation and the power transmission among all areas so that the cost function would be minimized while the load demand and constraints are still satisfied [30]-[36]. Mathematical modeling of multi area economic dispatch problem, decision variables, objective function and constraints are detailed in the following expression:

$$Min\ H(X) = \sum_{i=1}^{M} \sum_{j=1}^{Ngi} F_{ij}(P_{gij}) \quad (3)$$

where

$$F_{ij}(p_{gij}) = a_{ij} \times p_{gij}^2 + b_{ij} \times p_{gij} + c_{ij} + |e_{ij} \times \sin(f_{ij} \times (p_{gij\,min} - p_{gij}))| \quad (4)$$

Also,
$$Z = [\vec{p}_{g1}, \vec{p}_{g2}, \vec{p}_{g3}, \ldots, \vec{p}_{gM}, \vec{T}_1, \vec{T}_2, \ldots, \vec{T}_M] \quad (5)$$

$$[\vec{p}_{g1}, \vec{p}_{g2}, \vec{p}_{g3}, \ldots, \vec{p}_{gM}] = [Pg_{11}, Pg_{12}, Pg_{13}, \ldots, Pg_{1G_1}], [Pg_{21}, Pg_{22}, Pg_{23}, \ldots, Pg_{2G_2}], \ldots, [Pg_{M1}, Pg_{M2}, Pg_{M3}, \ldots, Pg_{MG_M}] \quad (6)$$

$$[\vec{T}_1, \vec{T}_2, \ldots, \vec{T}_M] = [T_{1,1}, T_{1,2}, \ldots, T_{1,M}], [T_{2,3}, T_{2,4}, \ldots, T_{2,M}], \ldots, [T_{M-1,M}] \quad (7)$$

In Some power plants multi types of fuel using as sources instead of one specific fuel for power generation. Thus, the coefficients of the cost function in each horizon are varied [15]. Therefore, by applying the valve point effect, prohibited operating zones and tie line capacity constraint, the multi-area economic dispatch lead to a complex, non-linear and non-convex problem. Consequently, a robust and effective optimization method is required to solve the problem [1].

### C. Constraint

#### 1. Power Generation Constraint

The power generation of each generator has a limitation as follows:
$$P_{gi\,min} \leq P_{gi} \leq P_{gi\,max} \quad (8)$$

#### 2. Load Balancing Constraint

Power generators should provide the total load demand and the transmission network losses. Hence, in multi area, the load demand in each area represents as the following expression:
$$P_{Gi} = P_{Di} + P_{Li} + \sum_{j=1, j \neq i}^{N} T_{ij} \quad i=1,2,\ldots,M \quad (9)$$

$P_{Li}$ is represented as the transmission network losses in the $i^{th}$ area as following [15]:
$$P_{Li} = \sum_{q=1}^{N_{gi}} \sum_{j=1}^{N_{gi}} P_{gij} B^i_{qj} P_{giq} + \sum_{j=1}^{N_{gi}} B^i_{0j} P_{gij} + B^i_{00} \quad (10)$$

#### 3. Prohibited Operating Zone Constraint

Because of some practical limitations which damages the plant or network instability, in some intervals each generator should not generate power. So, power generation in these intervals is prohibited. Fig (2), demonstrates the fuel curve of the power generators with two prohibited operating zones.

In addition, equation (11) shows the prohibited operating zone constraint as follows:

$$P_{gi} = \begin{cases} P_{gi\,min} \leq P_{gi} \leq P_{gi,l-1}^{Low} \\ P_{gi,l-1}^{Up} \leq P_{gi} \leq P_{gi,l}^{Low} \\ . \\ . \\ . \\ P_{gi,L_i}^{Up} \leq P_{gi} \leq P_{gimax} \end{cases} \quad l=2,3,\ldots,L_i \quad (11)$$

$i = 1,2,\ldots,N_g$

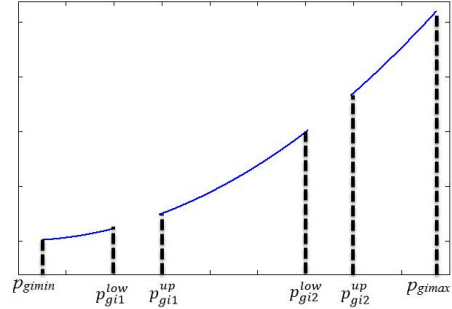

Fig. 2, The fuel curve of the power plant with two prohibited operating zone

#### 4. Tie Line Capacity Constraint

The transmission tie line capacity is one of the most important distinctive features constraint in MAED. Power exchange among areas should be between minimum and maximum capacity of the transmission line as follows:

$$-T_{i,j\,max} \leq T_{i,j} \leq T_{i,j\,max} \quad (12)$$

## III. ELECTRO SEARCH OPTIMIZATION ALGORITHM

In this paper, the Electro Search Optimization Algorithm (ESOA) is to overcome the nonlinearity of the problem. ESOA is a new meta-heuristic approach which is developed based on the theory of the electron movement around the nuclear atom. The main reason behind this approach is some significant advantages such as less mathematical calculation and no need to control parameters in comparison with other methods like particle swarm optimization (PSO), genetic algorithm (GA), simulated annealing (SA), hybrid PSO-GA algorithm (PGHA). ESOA is developed based on the following steps [24]:

*1) Atom spreading phase*

In the phase, the candidate solutions are spread all around the search space randomly.

*2) Orbital transition*

Based on the quantized energy concept, electrons around each nucleus effort to move to the greater orbit to achieve higher energy level.

$$e_i = N_i + (2 \times rand - 1)(1 - \frac{1}{n^2}) \times r, \quad n \in [2,5] \quad (13)$$

The new electrons in the greater level of energy are considered as the best solution ($e_{best}$) and utilized atom relocating in the next step [37-41].

*3) Nucleus relocation*

In nucleus relocation phase, the position of the new nucleus ($N_{new}$) is related to the energy stages of the atoms. As the result, the nucleus relocation is defined as follows:

$$\vec{D}_k = (\vec{e}_{best} - \vec{N}_{best}) + Re_k \otimes (\frac{1}{\vec{N}_{best}^2} - \frac{1}{\vec{N}_k^2}) \quad (14)$$

$$\vec{N}_{new,k} = \vec{N}_k + Ac_k \times \vec{D}_k \quad (15)$$

This step is continued for all nucleus and repositions of all atoms to find the global optimum point. According to the above equations, the convergence speed of the algorithm is related to the Rydberg's energy constant ($Re$) and accelerator coefficient ($Ac$) coefficients. These numbers are selected randomly for the first iteration while they will be updated in the next iteration using the following step information.

*4) Orbital-Tuner method*

Orbital-Tuner technique is utilized to update the Rydberg's energy and accelerator coefficients. This technique is proposed according to the cumulative normal density function. Therefore, instead of the center of gravity between two candidates, the center of mass is calculated.

$$Re_{k+1} = \frac{Re_k + (Re_{best} + \sum_{i=1}^{n} \frac{Re_i / f_{N_i/Re_i}}{1 / f_{N_i/Re_i}})}{2} \quad (16)$$

$$Ac_{k+1} = \frac{Ac_k + (Ac_{best} + \sum_{i=1}^{n} \frac{Ac_i / f_{N_i/Ac_i}}{1 / f_{N_i/Ac_i}})}{2} \quad (17)$$

In the ESOA, the initial numbers have no influence on the efficiency of the algorithm. This is due to these coefficients are controlled by self-tuning technique.

## IV. CASE STUDY AND RESULTS

In order to validate the effectiveness of the proposed algorithm, it is applied to two different cases as follows:
  *A. Six generators in two area*
  *B. Forty generators in four areas*

### A) Six generators in two areas

The selected test system includes six generation units in two different areas. Each area includes three generation units, in which the total load demand is 1263 MW. The load demand of the first area is 758.7 MW (60%) and for the second area is 505.2 MW (40%). The optimization problem includes prohibited operating zone, losses and the valve point effect. The capacity of the transmission line between two areas is 100 MW. Table I demonstrates the superiority of the proposed algorithm in comparison with other well-known optimization algorithm such as genetic algorithm (GA), and teaching-learning based optimization (TLBO).

Convergence speed is one of the significant factor for the heuristic algorithm. Figure 3 demonstrates the high convergence speed of the proposed algorithm.

Table I
Optimal Solution of case A

| Power Generation (MW) | ESOA | TLBO | GA |
|---|---|---|---|
| P11 | 499.94 | 500 | 500 |
| P12 | 200 | 200 | 200 |
| P13 | 150 | 150 | 150 |
| P21 | 199.87 | 204.3271 | 204.3341 |
| P22 | 146.63 | 154.7095 | 154.7048 |
| P23 | 75 | 67.5795 | 67.5770 |
| T12 | 87.68 | -- | -- |
| PL | 13.41 | 13.61 | 13.59 |
| Cost | 12210.66 | 12255.39 | 12255.42 |

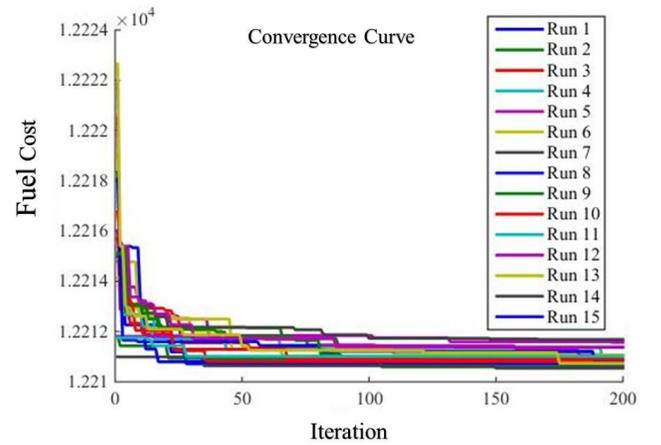

Fig. 3, Convergence curve for case

Based on the Fig. 3, the algorithm is optimized and reached the best answer in less than ten iterations.

### A) Forty generators in four areas

This system includes forty generators in four areas. This system is very complicated due to many nonlinear and nonconvex parameters. Fig. 4 shows eleven independent run for system B.

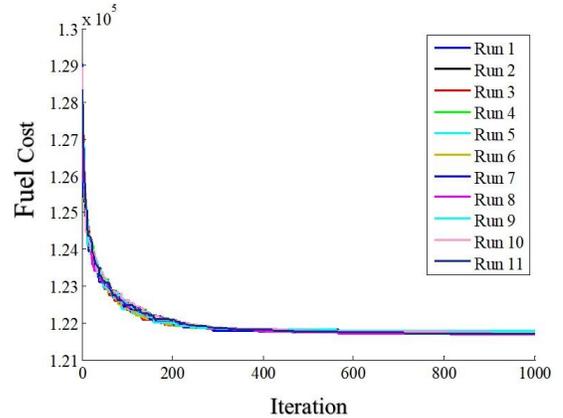

Fig. 4, Eleven run for system B

Table II compared the distribution of power generated obtained by ESOA and other methods.

Table II
Optimal Power generation for case B

| Power generation (MW) | ESOA | TLBO | GA |
|---|---|---|---|
| P11 | 110.8215 | 110.8791 | 111.5448 |
| P12 | 111.0259 | 112.955 | 111.7092 |
| P13 | 97.40176 | 97.4151 | 98.2429 |
| P14 | 179.7358 | 179.9466 | 179.8834 |
| P15 | 87.90663 | 89.4955 | 95.95 |
| P16 | 139.9999 | 139.8937 | 139.3533 |
| P17 | 259.5976 | 259.7338 | 259.3395 |
| P18 | 284.601 | 284.6387 | 285.3569 |
| P19 | 284.597 | 284.7414 | 284.9627 |
| P110 | 130 | 130.1151 | 130.2217 |
| P21 | 94.00712 | 168.8311 | 243.6005 |
| P22 | 94.01658 | 168.8214 | 95.389 |
| P23 | 304.5255 | 125.0623 | 214.5171 |
| P24 | 394.2784 | 394.2799 | 394.0808 |
| P25 | 394.2884 | 394.2529 | 394.2481 |
| P26 | 394.2787 | 484.0429 | 394.436 |
| P27 | 489.2967 | 489.284 | 489.9552 |
| P28 | 489.2897 | 489.2703 | 488.8885 |
| P29 | 534.7352 | 511.3347 | 511.4713 |
| P210 | 511.3035 | 511.4548 | 511.4125 |
| P31 | 523.2761 | 523.2816 | 523.2896 |
| P32 | 523.2831 | 523.4321 | 523.295 |
| P33 | 523.29 | 523.377 | 523.4129 |
| P34 | 523.2834 | 523.5974 | 523.4073 |
| P35 | 523.2865 | 523.5493 | 523.7703 |
| P36 | 523.2919 | 523.2773 | 523.5424 |
| P37 | 10.00002 | 10.1442 | 10.1621 |
| P38 | 10.0037 | 10.0248 | 10.1326 |
| P39 | 10 | 10.0862 | 10.6366 |
| P310 | 88.10562 | 88.2354 | 88.1189 |
| P41 | 190 | 189.919 | 161.222 |
| P42 | 189.9999 | 189.9718 | 189.5668 |
| P43 | 190 | 190 | 189.924 |
| P44 | 164.7963 | 164.8927 | 165.6621 |
| P45 | 164.8027 | 165.1343 | 165.4321 |
| P46 | 164.8005 | 165.2322 | 164.9868 |
| P47 | 89.14464 | 90.2758 | 109.8137 |
| P48 | 89.12672 | 109.9813 | 109.7935 |
| P49 | 102.5187 | 90.2019 | 90.1543 |
| P410 | 511.2834 | 458.9376 | 459.114 |
| T12 | 199.9866 | 185.5862 | 172.0652 |
| T13 | -7.82661 | -23.6686 | -36.3060 |
| T14 | -81.4729 | -47.1037 | -86.8070 |
| T23 | -199.994 | -183.0863 | -191.1128 |
| T24 | -99.9999 | -94.6933 | -98.8231 |
| T34 | -100 | -97.7497 | -45.0391 |
| Cost | 121694.384 | 121760.5 | 121794.8 |

Figure 5, compared the convergence curve EASO with GA as one of the well-known optimization algorithm. Based on the figure, GA algorithm is trapped in the local minimums, but EASO algorithm passed easily from the local minimums and reached the best solution.

One of the most important constraints in multi area economic dispatch problem is tie line capacity constraint. Fig 6, shows the iterative evolution of tie-line flows convergence in system B. According to the figure, the power transmission between areas converge very fast to the optimal value.

## V. CONCLUSION

Multi area economic dispatch is an advance form of economic dispatch. In multi area economic dispatch, in addition to the commonly constraint, the transmission constrains such as tile line capacity should be satisfied as well. In this paper, a new modified and effective algorithm known as the electro search optimization algorithm is proposed. In order to evaluate the ability and effectiveness of the algorithm, it is applied to MAED with different complexity. The results proved the high efficiency, performance, accuracy and robustness of the algorithm. Indeed, ESOA algorithm is simpler with less mathematical calculation, high speed, no need any control parameters and simple implementation for both constraint and unconstrained optimization problems. Moreover, by increasing the dimension and complexity of the problem, the performance of the algorithm is same.

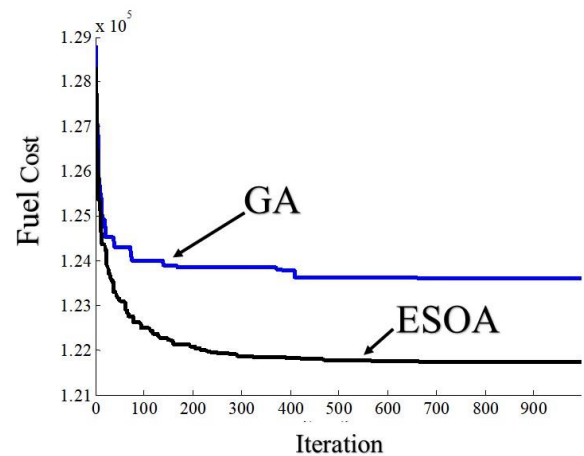

Fig 5. GA and EASO convergence curve for system B

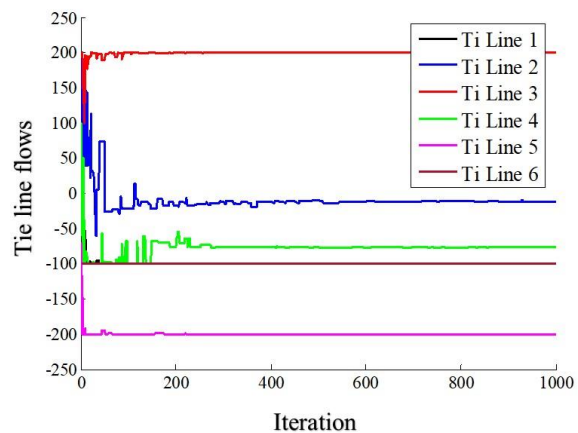

Fig 7. Iterative evolution of tie-line flows convergence in system B

REFERENCES

[1] M. Basu, "Teaching–learning-based optimization algorithm for multi-area economic dispatch," *Energy,* vol. 68, pp. 21-28, 2014.
[2] Dabbaghjamanesh, M., A. Moeini, M. Ashkaboosi, P. Khazaei, and K. Mirzapalangi. "High performance control of grid connected cascaded H-Bridge active rectifier based on type II-fuzzy logic controller with low frequency modulation technique." International Journal of Electrical and Computer Engineering 6, no. 2 (2016): 484.

[3] Rakhshan, Mohsen, Navid Vafamand, Mokhtar Shasadeghi, Morteza Dabbaghjamanesh, and Amirhossein Moeini. "Design of networked polynomial control systems with random delays: sum of squares approach." International Journal of Automation and Control 10, no. 1 (2016): 73-86

[4] Shahmaleki, P., M. Amiri, and M. Mahzoon. "Adaptive Neuro Fuzzy Control Design of a Solar Power Plant's Complete Oil Cycle." In ASME 2009 3rd International Conference on Energy Sustainability collocated with the Heat Transfer and InterPACK09 Conferences, pp. 625-633. American Society of Mechanical Engineers, 2009.

[5] Shahmaleki, Pourya, Mojtaba Mahzoon, Alireza Kazemi, and Mohammad Basiri. "Vision-based hierarchical fuzzy controller and real time results for a wheeled autonomous robot." In Motion Control. InTech, 2010.

[6] Ashkaboosi, Maryam, Seyed Mehdi Nourani, Peyman Khazaei, Morteza Dabbaghjamanesh, and Amirhossein Moeini. "An optimization technique based on profit of investment and market clearing in wind power systems." American Journal of Electrical and Electronic Engineering 4, no. 3 (2016): 85-91.

[7] Shahmaleki, Pourya, and Mojtaba Mahzoon. "GA modeling and ANFIS control design for a solar power plant." In American Control Conference (ACC), 2010, pp. 3530-3535. IEEE, 2010.

[8] Liu, Xian, and Hamzeh Davarikia. "Optimal Power Flow with Disjoint Prohibited Zones: New Formulation and Solutions." arXiv preprint arXiv:1805.03769 (2018).

[9] Ashkaboosi, Maryam, Farnoosh Ashkaboosi, and Seyed Mehdi Nourani. "The Interaction of Cybernetics and Contemporary Economic Graphic Art as" Interactive Graphics"." (2016).

[10] Z.-L. Gaing, "Particle swarm optimization to solving the economic dispatch considering the generator constraints," *Power Systems, IEEE Transactions on,* vol. 18, pp. 1187-1195, 2003.

[11] Ghaffari, Saeed, and M. Ashkaboosi. "Applying Hidden Markov M Recognition Based on C." (2016).

[12] W.-M. Lin, F.-S. Cheng, and M.-T. Tsay, "An improved tabu search for economic dispatch with multiple minima," *Power Systems, IEEE Transactions on,* vol. 17, pp. 108-112, 2002.

[13] Amin Sahba, Yilin Zhang, Marcus Hays, and Wei-Ming Lin. "A Real-Time Per-Thread IQ-Capping Technique for Simultaneous Multi-threading (SMT) Processors." In Information Technology: New Generations (ITNG), 2014 11th International Conference on, pp. 413-418. IEEE, 2014.

[14] Amin Sahba, and John J. Prevost. "Hypercube based clusters in Cloud Computing." In World Automation Congress (WAC), 2016, pp. 1-6. IEEE, 2016.

[15] Mahdi Bagheri, Mehdi Madani, Ramin Sahba, and Amin Sahba. "Real time object detection using a novel adaptive color thresholding method." In Proceedings of the 2011 international ACM workshop on Ubiquitous meta user interfaces, pp. 13-16. ACM, 2011.

[16] Amin Sahba, Ramin Sahba, and Wei-Ming Lin. "Improving IPC in simultaneous multi-threading (SMT) processors by capping IQ utilization according to dispatched memory instructions." In World Automation Congress (WAC), 2014, pp. 893-899. IEEE, 2014.

[17] Ramin Sahba, "Hashing for fast IP address lookup utilizing inter-key correlation." The University of Texas at San Antonio, 2013.

[18] J. Cai, Q. Li, L. Li, H. Peng, and Y. Yang, "A fuzzy adaptive chaotic ant swarm optimization for economic dispatch," *International Journal of Electrical Power & Energy Systems,* vol. 34, pp. 154-160, 2012.

[19] V. N. Dieu, W. Ongsakul, and J. Polprasert, "The augmented Lagrange Hopfield network for economic dispatch with multiple fuel options," *Mathematical and Computer Modelling,* vol. 57, pp. 30-39, 2013.

[20] A. L. Desell, E. C. McClelland, K. Tammar and P. r. V. Horne, "Transmission Constrained Production Cost Analysis in Power System Planning," in *IEEE Transactions on Power Apparatus and Systems*, vol. PAS-103, no. 8, pp. 2192-2198, Aug. 1984.

[21] M. Dabbaghjamanesh, S. Mehraeen, A. Kavousifard and M. A. Igder, "Effective scheduling operation of coordinated and uncoordinated wind-hydro and pumped-storage in generation units with modified JAYA algorithm," *2017 IEEE Industry Applications Society Annual Meeting*, Cincinnati, OH, 2017, pp. 1-8.

[22] Khazaei, Peyman, Morteza Dabbaghjamanesh, Ali Kalantarzadeh, and Hasan Mousavi. "Applying the modified TLBO algorithm to solve the unit commitment problem." In World Automation Congress (WAC), 2016, pp. 1-6. IEEE, 2016..

[23] L. Wang and C. Singh, "Reserve-Constrained Multiarea Environmental/Economic Dispatch Using Enhanced Particle Swarm Optimization," *2006 IEEE Systems and Information Engineering Design Symposium*, Charlottesville, VA, 2006, pp. 96-100.

[24] Khazaei, P., S. M. Modares, M. Dabbaghjamanesh, M. Almousa, and A. Moeini. "A high efficiency DC/DC boost converter for photovoltaic applications." International Journal of Soft Computing and Engineering (IJSCE) 6, no. 2 (2016): 2231-2307.

[25] M. Basu, "Artificial bee colony optimization for multi-area economic dispatch," *International Journal of Electrical Power & Energy Systems,* vol. 49, pp. 181-187, 2013.

[26] Sun, C., and V. Jahangiri. "Bi-directional vibration control of offshore wind turbines using a 3D pendulum tuned mass damper." Mechanical Systems and Signal Processing 105 (2018): 338-360.

[27] Jahangiri, Vahid, and Mir Mohammad Ettefagh. "Multibody Dynamics of a Floating Wind Turbine Considering the Flexibility Between Nacelle and Tower." International Journal of Structural Stability and Dynamics (2017): 1850085.

[28] Jahangiri, Vahid, Hadi Mirab, Reza Fathi, and Mir Mohammad Ettefagh. "TLP Structural Health Monitoring Based on Vibration Signal of Energy Harvesting System." Latin American Journal of Solids and Structures 13, no. 5 (2016): 897-915.

[29] Mirab, Hadi, Reza Fathi, Vahid Jahangiri, Mir Mohammad Ettefagh, and Reza Hassannejad. "Energy harvesting from sea waves with consideration of airy and JONSWAP theory and optimization of energy harvester parameters." Journal of Marine Science and Application 14, no. 4 (2015): 440-449.

[30] Dabbaghjamanesh, Morteza, Abdollah Kavousi-Fard, and Shahab Mehraeen. "Effective Scheduling of Reconfigurable Microgrids with Dynamic Thermal Line Rating." *IEEE Transactions on Industrial Electronics* (2018).

[31] Dabbaghjamanesh, Morteza, Shahab Mehraeen, Abdollah Kavousi Fard, and Farzad Ferdowsi. "A New Efficient Stochastic Energy Management Technique for Interconnected AC Microgrids." *arXiv preprint arXiv:1803.03320* (2018).

[32] Ghaffari, Saeed, and Maryam Ashkaboosi. "Applying Hidden Markov Model Baby Cry Signal Recognition Based on Cybernetic Theory." *IJEIR* 5, no. 3 (2016): 243-247.

[33] H. Davarikia, F. Znidi, M. R. Aghamohammadi, K. Iqbal. Identification of coherent groups of generators based on synchronization coefficient. 2016 IEEE Power and Energy Society General Meeting (PESGM). 2016:1-5.

[34] F. Znidi, H. Davarikia and K. Iqbal, "Modularity clustering based detection of coherent groups of generators with generator integrity indices," in 2017 IEEE Power & Energy Society General Meeting, 2017, pp. 1-5.

[35] H. Davarikia, "Investment Plan Against Malicious Attacks on Power Networks: Multilevel Game-Theoretic Models with Shared Cognition." PhD diss., University of Arkansas at Little Rock, 2017.

[36] H. Davarikia, M. Barati, F. Znidi, K. Iqbal, "Real-Time Integrity Indices in Power Grid: A Synchronization Coefficient Based Clustering Approach," arXiv Preprint arXiv:1804.02793, 2018.

[37] Farshid Sahba et al., "Wireless Sensors and RFID in Garden Automation", International Journal of Computer and Electronics Research, vol. 3, no. 4, August 2014

[38] Farshid Sahba, Zahra Nourani, "The diagnosis of lumbar disc disorder by MR image processing and data mining", presented at the 2016 World Automation Congress, 2016

[39] Mahdi Keshavarz Bahaghighat, Farshid Sahba, Ehsan Tehrani, "Textdependent Speaker Recognition by combination of LBG VQ and DTW for persian language", Intermtiornl Jounral of Computer applications (0975 - 8887) volume51-no.16, August 2012

[40] Farshid Sahba, "Museum automation with RFID", Proceedings of the 2014 World Automation Congress (WAC), pp. 19-22, Aug 2014

[41] Mina Yazdandoost, Ali Yazdandoost, Fakhri Akhoonili, Farshid Sahba, "Smart tractors in pistachio orchards equipped with RFID", presented at the 2016 World Automation Congress, 2016.